\documentclass[superscriptaddress,prl,nofootinbib,showpacs,preprintnumbers]{revtex4}

\usepackage{amsfonts}
\usepackage{mathrsfs}
\usepackage{amssymb}
\usepackage{amsmath}
\usepackage{graphicx,epsfig}

\def\bwt{\begin{widetext}}

\def\ewt{\end{widetext}}

\def\be{\begin{equation}}

\def\ee{\end{equation}}

\def\bea{\begin{eqnarray}}

\def\eea{\end{eqnarray}}

\def\bean{\begin{eqnarray*}}

\def\eean{\end{eqnarray*}}

\def\bary{\begin{array}}

\def\eary{\end{array}}

\def\bit{\begin{itemize}}

\def\eit{\end{itemize}}

\begin{document}

\title{Time Delays of Strings in D-particle Backgrounds and Vacuum Refractive Indices}

\author{Tianjun Li}

\affiliation{George P. and Cynthia W. Mitchell Institute for
Fundamental Physics, Texas A$\&$M University, College Station, TX
77843, USA }

\affiliation{ Key Laboratory of Frontiers in Theoretical Physics,
  Institute of Theoretical Physics, Chinese Academy of Sciences,
  Beijing 100190, P. R. China}

\author{Nick E. Mavromatos}

\affiliation{Theoretical Physics, Department of Physics, King's College
London, Strand, London WC2R 2LS, U.K.}

\author{Dimitri V. Nanopoulos}

\affiliation{George P. and Cynthia W. Mitchell Institute for
Fundamental Physics,
 Texas A$\&$M University, College Station, TX 77843, USA }

\affiliation{Astroparticle Physics Group,
Houston Advanced Research Center (HARC),
Mitchell Campus, Woodlands, TX 77381, USA}

\affiliation{Academy of Athens, Division of Natural Sciences,
 28 Panepistimiou Avenue, Athens 10679, Greece }

\author{Dan Xie}

\affiliation{George P. and Cynthia W. Mitchell Institute for
Fundamental Physics, Texas A$\&$M University, College Station, TX
77843, USA }



\begin{abstract}

Using standard techniques in string/D-brane scattering amplitude computations,
we evaluate the scattering of open strings off D-particles in brane world scenarios.
The D-particles are viewed as D3 branes wrapped up around three cycles,
and their embedding in brane worlds constitutes a case of intersecting branes,
among which strings are stretched, representing various types of excitations of
the Standard Model (SM) particles in the low-energy limit.
Our analysis, reveals interesting and novel selection rules for
the resulting causal time delays, proportional to the energy of the incident
matter state, from the processes of splitting, capture and re-emission of the
latter by the D-particles.
In particular, we show that there are relatively large time delays only for
excitations that belong to the Cartan subalgebra of the SM gauge group,
which notably includes photons. We discuss the possible relevance of these
results to the models of space time foam predicting a non-trivial vacuum refractive
index for photons and the associated cosmic $\gamma$-ray phenomenology. In particular,
we demonstrate how low-string-scale models can be falsified already in this context, by current
astrophysical observations of cosmic photons, like those of MAGIC, H.E.S.S. and FERMI Telescopes.
In this way, such observations serve as a way to discriminate low- from high-string scale models.

\end{abstract}


\preprint{ACT-04-09, MIFP-09-10}

\maketitle

\section{Introduction: Lorentz-Invariance-Violating String-Foam Backgrounds }

The advent of the new generation of ground-based {\v C}erenkov
or satellite $\gamma$-ray telescopes has inaugurated a new
era in $\gamma$-ray astronomy. In particular, these
instruments may also be used to probe fundamental physics,
for example, a possible energy-dependent vacuum refractive index for photon
due to quantum-gravitational effects in space-time foam~\cite{robust,emnnewuncert}.
MAGIC~\cite{MAGIC2}, HESS~\cite{hessnew} and Fermi~\cite{grbglast} Collaborations have
reported time-lags in the arrival times of high-energy photons, as compared with
photons of lower energies. The most conservative interpretations of
such time-lags are that they are due to emission mechanisms at the sources,
which are still largely unknown at present. However, such delays might also
 be the hints for the energy-dependent vacuum refractive index.
Assuming that the refractive index $n$ depends linearly on the $\gamma$-ray
energy $E_{\gamma}$,
{\it i.e.}, $n \sim 1 + E_{\gamma}/M_{\rm QG}$ where
$M_{\rm QG}$ is the quantum gravity scale, it was shown that
the time delays observed by the MAGIC~\cite{MAGIC2}, HESS~\cite{hessnew}
and Fermi~\cite{grbglast} Collaborations are compatible with
each other~\cite{emnnewuncert} for $M_{\rm QG}$ around $10^{18}$ GeV.

It is well known that many theories of quantum gravity predict
non-trivial vacuum refractive indices, varying linearly
with the energy scale of the photons and with the distance of the source.
However,  there are several
stringent restrictions coming from other independent tests of Lorentz symmetry
that must be taken into account~\cite{emnnewuncert}.
To survive these stringent bounds on Lorentz invariance
violation imposed by the plethora of the current astrophysical experiments,
any model of quantum-gravity predicting non-trivial refractive indices
should be characterised by the following features:
\textbf{{(i)}} Photons are \emph{stable} ({\it i.e.}
 do \emph{not} decay)~\cite{sigl} but should exhibit a
modified \emph{subluminal} dispersion relation with Lorentz-violating corrections that
should grow \emph{linearly} with $E_{\gamma}/M_{\rm QG}$ where $M_{\rm QG}$ is
close to the Planck scale;
\textbf{{(ii)}} The medium should not refract electrons, so as to avoid the
synchrotron-radiation constraints~\cite{crab,ems};
\textbf{{(iii)}}  The coupling of the photons to the medium must be independent of photon
polarization, so as
to avoid birefringence, thus avoiding the stringent pertinent
constraints~\cite{uv,grb,crab2,macio};
\textbf{{(iv)}} The formalism of a local effective field theory lagrangian
in an effectively flat space-time, including higher-derivative local
interaction terms to produce a refractive index~\cite{myers}, breaks down.
This would be signalled by quantum fluctuations in the total energy in
particle interactions~\cite{emngzk}, due to the presence of
 a quantum-gravitational `environment'.
\begin{figure}[th]
\centering
\includegraphics[width=7.5cm]{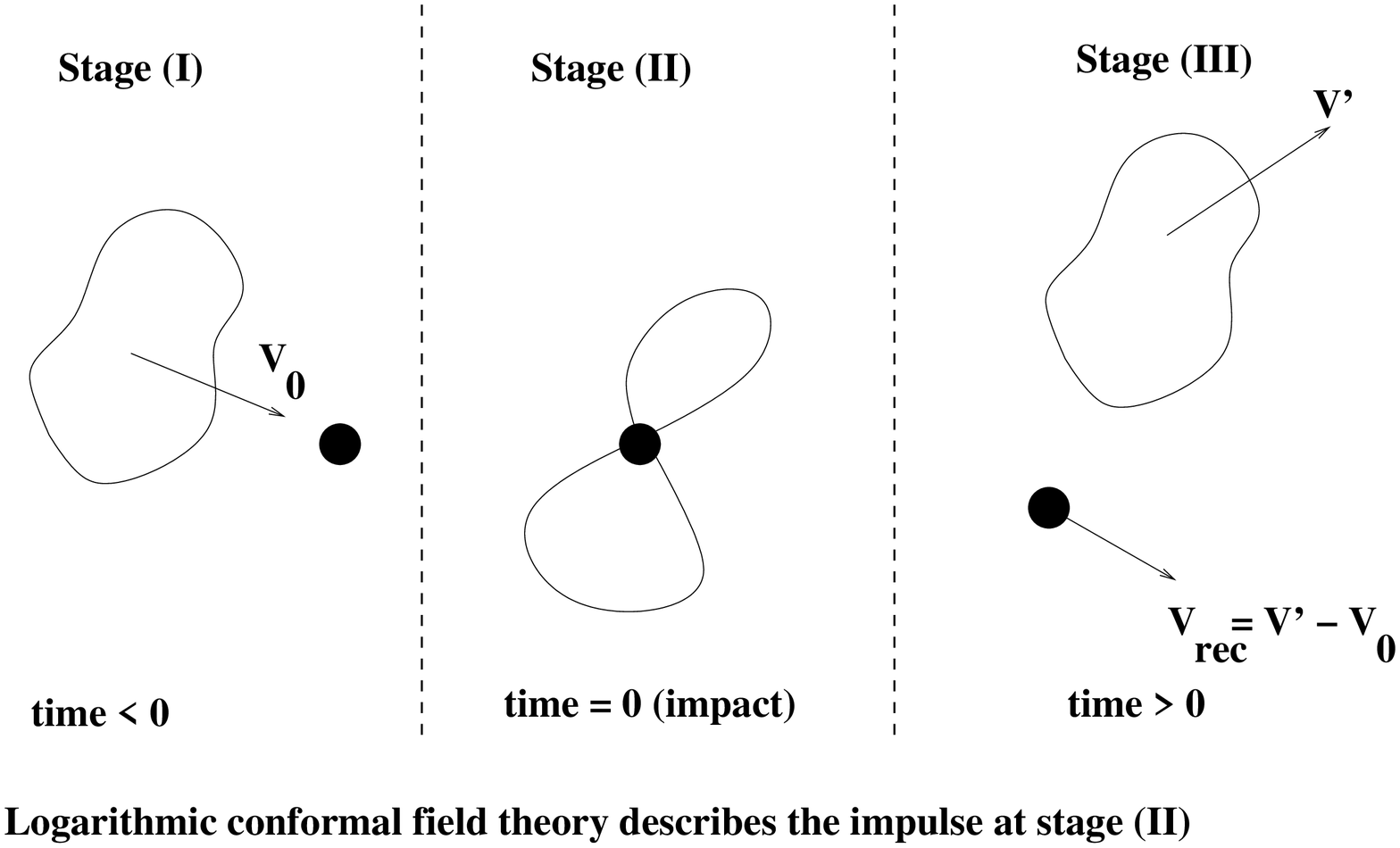} \hfill
\includegraphics[width=7.5cm]{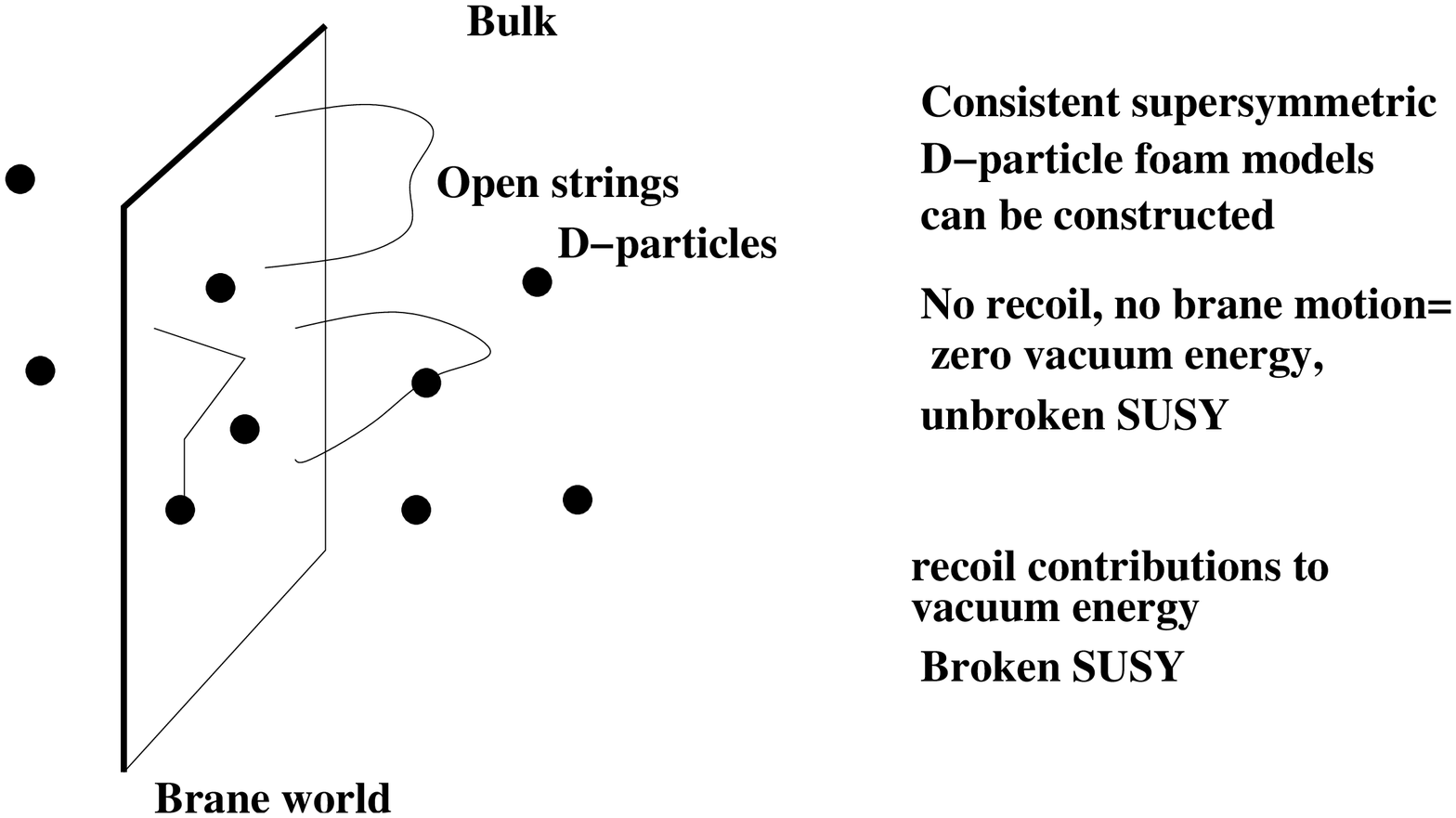} \caption{Schematic
representation of a D-particle foam. The figure indicates also the capture/recoil
process of a string state by a D-particle defect for closed (left) and open
(right) string states, in the presence of D-brane world. The presence of a
D-brane is essential due to gauge flux conservation, since an isolated
D-particle cannot exist. The intermediate composite state at $t=0$ has
a life time $\Delta t = \alpha ' p^0$, where $p^0$ is the energy of the incident
string state, which is compatible with the space-time string uncertainty principles~\cite{yoneya},
as explained in \cite{emnnewuncert}. When the (open) string sate represents a photon, then the (causal) delay $\Delta t$
will lead to \emph{non-trivial optical properties} (refractive index) for this space time, independent though of the photon polarization, and hence such an effect will not imply birefringence. Moreover, D-particle recoil during this splitting/capture process is responsible for the distortion of the surrounding space time during the scattering, and this leads to induced metrics depending on both coordinates and momenta of the string state. This results on modified dispersion relations for the open string propagation in such a background, which adds to the above-mentioned non-trivial
optical properties of the D-particle foam.}%
\label{fig:recoil}%
\end{figure}
The only known model with all these properties has been suggested by J. Ellis and
two of us~\cite{horizons,emnw,ems} within the framework of string/brane theory,
based on a stringy analogue of the interaction of a photon with internal
degrees of freedom in a conventional medium. The space-time foam is modelled
as a gas of point-like D-brane defects (D-particles) in the bulk space-time of
a higher-dimensional cosmology where the observable Universe is a D3-brane (c.f. fig.~\ref{fig:recoil}).
Within this class of D-foam models, a refractive index for photon propagation
in vacuo was re-derived by using a detailed modelling of the interaction of
an open string, representing a photon, with a D-particle~\cite{emnnewuncert}.

Indeed, in such a model, photons are represented as electrically neutral open string states, which are captured by D-particles during topologically non-trivial scattering processes, in which an incident open string, with momentum $p^0$ splits into constituent intermediate string states, stretched between the D-particle and the D3 brane worlds. The life time of such intermediate states can be calculated, as in \cite{sussk}, by means of computing the (causal) time delays required for the first of a series of re-emitted outgoing waves, with attenuating amplitudes. It is found proportional to the incident energy of the photon,
\begin{equation}
\Delta t \sim \alpha '\,p^0~,
\label{timedelay}
\end{equation}
Equivalently, this time delay is the time taken for an intermediate string state, stretched between the D-particle and the D3 brane to grow from zero size to a maximum one and back to zero, with the end of the string attached to the D3 brane moving with the speed of light~\cite{emnnewuncert,sussk}. This is a purely stringy effect, not existing in any local quantum field theory limit, and is compatible with the string space-time uncertainty relations~\cite{yoneya} for this case.
The result (\ref{timedelay}) admits an interpretation~\cite{emnnewuncert} in terms of a non-trivial \emph{subluminal} refractive index of the D-particle foam, which scales linearly with the photon energy.
Interestingly, the D-particle foam looks transparent to charged particles
such as electrons due to charge conservation, and then avoids the stringent
constraints coming from synchrotron radiation of the Crab
Nebula~\cite{crab,crab2,ems}.

In addition to this leading refractive index effect, obtained form the causal outgoing waves during the scattering of strings with D-particles, there are corrections induced by the recoil of the D-particle itself, which contribute to space-time distortions that we now proceed to discuss.
To this end we note that, from a world-sheet view point, the presence of $D$-particle recoil may be represented by adding to a fixed-point (conformal) $\sigma$-model action, the following deformation~\cite{mavronew}:
\begin{equation}
\mathcal{V}_{\rm{D}}^{imp}=\frac{1}{2\pi\alpha '}
\sum_{i=1}^{d}\int_{\partial D}d\tau\,u_{i}%
X^{0}\Theta\left(  X^{0}\right)  \partial_{n}X^{i}. \label{fullrec}%
\end{equation}
where $d$ in the sum denotes the appropriate  number of spatial target-space dimensions.
For a recoiling D-particle confined on a D3 brane, $d=3$.

On writing the boundary recoil/capture operator (\ref{fullrec}) as a total derivative over the bulk of the world-sheet, by means of the two-dimensional version of Stokes theorem we have (omitting from now on the explicit summation over repeated $i$-index, which is understood to be over the spatial indices of the D3-brane world):
\begin{eqnarray}\label{stokes}
&& \mathcal{V}_{\rm{D}}^{imp}=\frac{1}{2\pi\alpha '}
\int_{D}d^{2}z\,\epsilon_{\alpha\beta} \partial^\beta
\left(  \left[  u_{i}X^{0}\right]  \Theta\left(  X^{0}\right)  \partial^{\alpha}X^{i}\right) = \nonumber \\
&& \frac{1}{4\pi\alpha '}\int_{D}d^{2}z\, (2u_{i})\,\epsilon_{\alpha\beta}
(\partial^{\beta}X^{0})\Bigg[ \Theta_\varepsilon \left(X^{0}\right) + X^0 \delta_\varepsilon \left(  X^{0}\right) \Bigg] \partial
^{\alpha}X^{i}
\end{eqnarray}
where $\delta_\varepsilon (X^0)$ is an $\varepsilon$-regularised $\delta$-function.
This is equivalent to a deformation describing an open string propagating in an antisymmetric  $B_{\mu\nu}$-background corresponding to an external constant in target-space ``electric'' field,
\begin{equation}
B_{0i}\sim u_i ~, \quad B_{ij}=0~,
\label{constelectric}
\end{equation}
where the $X^0\delta (X^0)$ terms in the argument of the electric field yield vanishing contributions in the large time limit $\varepsilon \to 0$, and hence are ignored from now on.

To discuss the space time effects of a recoiling D-particle on an open  string state propagating on a D3 brane world, which is of interest to us here, we should consider a $\sigma$-model, in the presence of the B-field, which leads to mixed-type boundary conditions for open strings on the boundary $\partial \mathcal{D}$ of world-sheet surfaces with the topology of a disc:
\begin{equation}
      g_{\mu\nu}\partial_n X^\nu + B_{\mu\nu}\partial_\tau X^\nu |_{\partial \mathcal{D}} = 0~,
\label{bc}
\end{equation}
with $B$ given by (\ref{constelectric}). Absence of a recoil-velocity $u_i$-field leads to the usual Neumann boundary conditions, while the limit where $g_{\mu\nu} \to 0$, with $u_i \ne 0$, leads to Dirichlet boundary conditions.

Considering commutation relations among the coordinates of the first quantised $\sigma$-model in the above background, one also obtains a non-commutative space-time relation~\cite{sussk}~\footnote{In contrast, in the case of strings in a constant magnetic field~\cite{seibergwitten} (corresponding to B-fields of the form $B_{ij} \ne 0$, $B_{0i~}=0$), the non commutativity is only between spatial target-space coordinates}. The pertinent non commutativity refers to spatial coordinates along the direction of the electric field, and is expressed in the form
\begin{equation}
[ X^1, t ] = i \theta^{10} ~, \qquad \theta^{01} (= - \theta^{10}) \equiv \theta =  \frac{1}{u_{\rm c}}\frac{\tilde u}{1 - \tilde{u}^2}
\label{stnc}
\end{equation}
where $t$ is the target time, and we assume for simplicity and concreteness recoil along the spatial $X^1$ direction. Thus, the induced non commutativity is consistent with the breaking of the Lorentz symmetry of the ground state by the D-particle recoil. The quantity $\tilde{u}_i \equiv \frac{u_i}{u_{\rm c}}$ and  $u_{\rm c} = \frac{1}{2\pi \alpha '}$ is the Born-Infeld \emph{critical} field.
Notice that the presence of the critical ``electric'' field is associated with a singularity of both the effective metric and the non commutativity parameter, while, as we shall discuss below (\ref{effstringcoupl}) there is also an effective string coupling, which  vanishes in that limit. This reflects the \emph{destabilization of the vacuum} when the ``electric'' field intensity approaches the \emph{critical value}, which was noted in \cite{burgess}. Since in our D-particle foam case, the r\^ole of the `electric' field is played by the recoil velocity of the
D-particle defect, the critical field corresponds to the relativistic speed of light, in accordance with special relativistic kinematics, which is respected in string theory, by construction.

The space-time uncertainty relations (\ref{stnc}) are consistent with the corresponding space-time string uncertainty principle~\cite{yoneya}
\begin{equation}
   \Delta X \Delta t \ge \alpha '
\label{stringyunc}
\end{equation}
Of crucial interest in our case is the form of the induced open-string \emph{effective target-space-time metric}. As discussed in detail in refs.~\cite{sussk,seibergwitten}, to find it,
one should consider the world-sheet propagator on the disc $\langle X^\mu(z,{\overline z})X^\nu(0,0)\rangle$, with the boundary conditions (\ref{bc}).
Upon using a conformal mapping of the disc onto the upper half plane
with the real axis (parametrised by $\tau \in R$) as its boundary~\cite{seibergwitten},
one then obtains:
\begin{equation}
     \langle X^\mu(\tau)X^\nu(0)\rangle = -\alpha ' g^{\mu\nu}_{\rm open,~electric}{\rm ln}\tau^2 + i\frac{\theta^{\mu\nu}}{2}\epsilon(\tau)
\label{propdisc}
\end{equation}
with the non-commutative parameters  $\theta^{\mu\nu}$ given by by (\ref{stnc}),
and the effective open-string metric, due to the presence of the recoil-velocity field $\vec{u}$, whose direction breaks target-space Lorentz invariance, by:
\begin{eqnarray}
           g_{\mu\nu}^{\rm open,electric} &=& \left(1 - {\tilde u}_i^2\right)\eta_{\mu\nu}~, \qquad \mu,\nu = 0,1 \nonumber \\
           g_{\mu\nu}^{\rm open,electric} &=& \eta_{\mu\nu}~, \mu,\nu ={\rm all~other~values}~,
\label{opsmetric}
\end{eqnarray}
where, for concreteness and simplicity, we consider a frame of reference where the matter particle
has momentum only across the spatial direction $X^1$, \emph{i.e.} $0 \ne k_1 \equiv k \parallel u_1~, k_2=k_3 =0$.
Moreover, there is a modified effective string coupling~\cite{seibergwitten,sussk}:
\begin{equation}
   g_s^{\rm eff} = g_s \left(1 - \tilde{u}^2\right)^{1/2}
\label{effstringcoupl}
\end{equation}
The fact that the metric in our recoil case depends on momentum transfer variables, implies that D-particle recoil induces Finsler-type metrics~\cite{finsler}, \emph{i.e}. metric functions that depend on phase-space coordinates, that is space-time and momentum coordinates.

The induced metric (\ref{opsmetric}), will affect the dispersion relations of the photon state,
by means of $k^\mu k^\nu g{\mu\nu}^{\rm open,electric} = 0$. However, because the corrections on the recoil velocity $u_i$ are quadratic, such modifications will be suppressed by the square of the D-particle mass scale. The presence of the D-particle recoil velocity will affect the induced time delays (\ref{timedelay})
 by higher-order corrections of the form, as follows by direct analogy of our case with that of open strings in  a constant electric field~\cite{sussk}:
 \begin{equation}\label{timerecoil}
 \Delta t_{\rm with~D-foam~recoil~velocity} = \alpha ' \,\frac{p^0}{1 - {\tilde u}_i^2}~.
\end{equation}
As we mentioned previously, one observes that the D-particle recoil effects are quadratically suppressed by the D-particle mass scales, since ${\tilde u}_i \propto g_s \Delta k_i /M_s$, with $\Delta k_i$ the relevant string-state momentum transfer.

Some important remark is in order here, to avoid some confusion regarding testing the predictions of our model with experiment. As we hope becomes clear from \cite{emnnewuncert} and our discussion above, the primary time delay (\ref{timedelay}), is derived in the limit of infinite D-particle mass. The presence of D-particles, of course, breaks the Poincare invariance of the bulk vacuum, and their recoil the associated Lorentz invariance of the D3-brane world. In this sense, it is to be expected that non-trivial refractive indices characterise our D-particle foam background, even in the limit of infinite D-particle mass, where recoil motion of the D-particles is ignored. Having said that, it is also clear from the above analysis, and that of ref.~\cite{emnnewuncert}, that the calculation of time delays (\ref{timedelay}) is based on computing the formation of an intermediate string and \emph{not} point-like state, of a \emph{finite} extent. In this sense, such effects \emph{cannot} be simply \emph{represented} in terms of an effective \emph{local} low-energy \emph{lagrangian} formalism, in which one adds higher-derivative, higher-dimension non-renormalizable operators in a flat space-time lagrangian.
On the other hand, the effects of the induced Finsler metric, due to D-particle recoil, as well as their non-commutative contributions (\ref{stnc})~\footnote{For instance, on averaging ($\ll \dots \gg$) the non-commutativity relation (\ref{stnc}) over populations of (quantum fluctuating) D-particles in the foam, and assuming an approximately constant $\ll u_i \gg $, Eq.~\ref{stnc} becomes of the form $[x^\mu\, , \, x^\nu] = i\theta^{\mu\nu}$,
with $\theta^{\mu\nu} \simeq $~const. The latter relation may be interpreted~\cite{carroll} as implying a Lorentz violating situation, leading, in the continuum low-energy field theory limit, to higher-order derivative terms in a local effective lagrangian, which, for the case of photons we are interested in, would include terms of the form:
$$ {\cal L}^{EFT-NC} \ni -\frac{1}{4}F_{\mu\nu}F^{\mu\nu} - \frac{1}{2}\theta^{\alpha\beta}F_{\alpha\mu}F_{\beta\nu}F^{\mu\nu} +
\frac{1}{8}\theta^{\alpha\beta}F_{\alpha\beta}F_{\mu\nu}F^{\mu\nu} + \dots $$
The crucial qualitative difference in our case is the Finsler type of the non-commutative space-time parameter, $\theta^{0i}$ which in general depends on momenta. Nevertheless, such links of our model with effective field theories of non-commutative space times imply that the D-particle recoil parameters can be constrained also by means of the same experiments that are used to constrain such non commutative field theories~\cite{szabonc}. However, in our cases of D-particle foam, the most physically interesting case are that of isotropic, Lorentz-invariant on average, Gaussian stochastically fluctuating foam, for which the average recoil velocity of D-particles vanishes, $\ll u_i \gg =0$, and only their fluctuations are non trivial, $\ll u_i u_j \gg = \sigma^2 \delta_{ij}$, \emph{etc}. In such a case, non commutativity is much harder to detect via the methods discussed in \cite{szabonc,carroll}, although the independent phenomenon of time delays (\ref{timedelay}), (\ref{timerecoil}), which is not linked to effective field theories, is much easier to falsify in high-energy gamma ray astrophysics observations, like those of \cite{MAGIC2}.}, may admit such an effective field theory interpretation. Hence, the best experiments to test our model are the ones in which simply one computes time delays in the arrival times of photons (or electrically neutral particles) and checks their dependence on their energies. This has been done in \cite{emnnewuncert}. In this respect, we should stress that the analysis in \cite{MAGIC2}, where a reproduction of the peak has been used at an individual photon level, assuming linearly modified dispersion relations for photons, does not capture correctly the spirit of the present model, in view of these comments, in the sense that the latter analysis presupposes the existence of an underlying local effective lagrangian. It is for the same reason, namely the de-association of the effect (\ref{timerecoil}) from the effective field theory formalism, that ultrahigh energy cosmic rays constraints analyses~\cite{sigl} also do not apply to our stringy D-foam framework.

In the model of \cite{emnnewuncert} the D-particles are point-like D0-branes, which are admitted in type IIA string theories. In this letter we consider an extension of these ideas in the phenomenologically
relevant Type IIB string theory with in which the r\^ole of D-particles is played by
D3-branes appropriately wrapped around three cycles, while the D3 brane worlds are provided by
D7-branes, appropriately compactified to three large spatial dimensions.
We derive the vacuum refractive index for the photon, and find that it
depends linearly on the photon energy. In addition, in this model, in contrast to that of \cite{emnnewuncert}, there can be a time-delay for electron, which however, as we shall show, can be small enough to satisfy the  stringent constraints coming from synchrotron radiation
of the Crab Nebula~\cite{crab,crab2,ems}. We shall derive the time delays in the spirit of \cite{emnnewuncert}, by considering scattering amplitude first in the limit where the recoil of the D-particle is ignored.
As we shall see, a formula analogous to (\ref{timedelay}) will be derived.
The inclusion of wrapped-up-D3-brane/D-particle recoil corrections parallels that of D0-brane above,
and will not be repeated. The result is similar to (\ref{timerecoil}), with the D3-brane recoil to amount to quadratic and higher-order suppression by the effective ``D-particle'' mass.

\section{A type IIB String/Brane Foam Background and Causal Time Delays}

Let us consider the Type IIB string theory with D3-branes
and D7-branes where the D3-branes are inside the D7-branes.
The D3-branes wrap a three-cycle, and the D7-branes wrap a four-cycle.
Thus, the D3-branes can be considered as point
particles in the Universe, {\it i.e.}, the D-particles, while
the Standard Model (SM)
 particles are on the world-volume of the D7-branes.

For simplicity, we assume that the three internal space dimensions,
which the D3-brane wraps around, are cycles $S^1\times S^1\times S^1$, and we denote
the radius of the $i-$th cycle $S^1$ as $R_i$.
The mass of the D3-brane is~\cite{Johnson:2000ch}
\begin{equation}
M_{\rm D3} ~=~ {{R_1 R_2 R_3}\over {g_s  \ell_s^4}}~,~\,
\label{D3-Mass}
\end{equation}
whereby $g_s$ is the string coupling, and
 $\ell_s$ is the string length, {\it i.e.},
 the square root of the Regge
slope $\sqrt{\alpha '}$. If $R_i \le  \ell_s$, we can perform a
T-duality transformation along the $i-$th $S^1$ as follows
\begin{equation}
R_i \longrightarrow {{\ell_s^2}\over {R_i}}~,~\,
\end{equation}
and then we obtain $R_i \ge  \ell_s$. In string theories with compact internal space dimension(s),
there are Kaluza-Klein (KK) modes and string winding modes. Under
T-duality, the KK modes and winding modes are interchanged.
In particular, these two theories are physically identical~\cite{Johnson:2000ch}.
Therefore, without loss of generality, we can  assume
$R_i \ge  \ell_s$. Choosing $R_i = 10  \ell_s$ and $g_s \sim 0.5$,
we obtain $M_{\rm D3} \sim 2000/\ell_s$, and then the mass of the
D3-brane can be much larger than the string scale.
Therefore, anticipating the result (\ref{timerecoil}),
we shall ignore to a first approximation
the recoil of the D-particles. Their inclusion (as small,
 perturbative corrections) is straightforward,
 according to the discussion in the previous section,
and we shall come back briefly to this issue at the
end of the article. In this setting,
for the particles (called ND particles) arising
from the open strings between the D7-branes and D3-branes
which satisfy the Neumann (N) and Dirichlet (D) boundary
conditions respectively on the D7-branes and D3-branes, their
 gauge couplings with the gauge fields on the D7-branes are
\begin{equation}
\frac{1}{g_{37}^2} = \frac{V}{g_7^2}~,~\,
 \label{coupl}
\end{equation}
where $g_7$ are the gauge couplings on the D7-branes, and
$V$ denotes the volume of the extra four spatial dimensions of the D7 branes
transverse to the D3-branes~\cite{kutasov}.
Because the Minkowski space dimensions are non-compact,
$V$ is infinity and then $g_{37}$ is zero. Thus, the SM particles have no
interactions with the ND particles on the
D3-brane or D-particle.

To have non-trivial interactions between the particles on D7-branes
and the ND particles, we consider
a D3-brane \emph{foam}, {\it i.e.}, the D3-branes are distributed \emph{uniformly}
in the Universe. We assume that $V_{A3}$ is the average three-dimensional
volume that has a D3-brane in the Minkowski space dimensions, and $R'$ is the
radius for the fourth space dimension transverse to the D3-branes.
In addition, in the conformal field theory description, a D-brane is
an object with a well defined position. While in the string field theory,
a D-brane is a fat object with thickness of the order of the string
scale. In particular, the widths of the D-brane along the transverse
dimensions are about $1.55\ell_s$, as follows from an analysis of
the tachyonic lump solution in the string field theory
which may be considered as a D-brane~\cite{Moeller:2000jy}.
Thus, our ansatz for the gauge couplings between
 the gauge fields on the D7-branes and the ND particles is
\begin{equation}
\frac{1}{g_{37}^2} = \frac{V_{A3} R'}{(1.55\ell_s)^4}
\frac{\ell_s^4}{g_7^2} = \frac{V_{A3} R'}{(1.55)^4} \frac{1}{g_{7}^2}~.~\,
 \label{coupl-N}
\end{equation}

\begin{figure}[ht]
\centering
\includegraphics[width=0.6\textwidth]{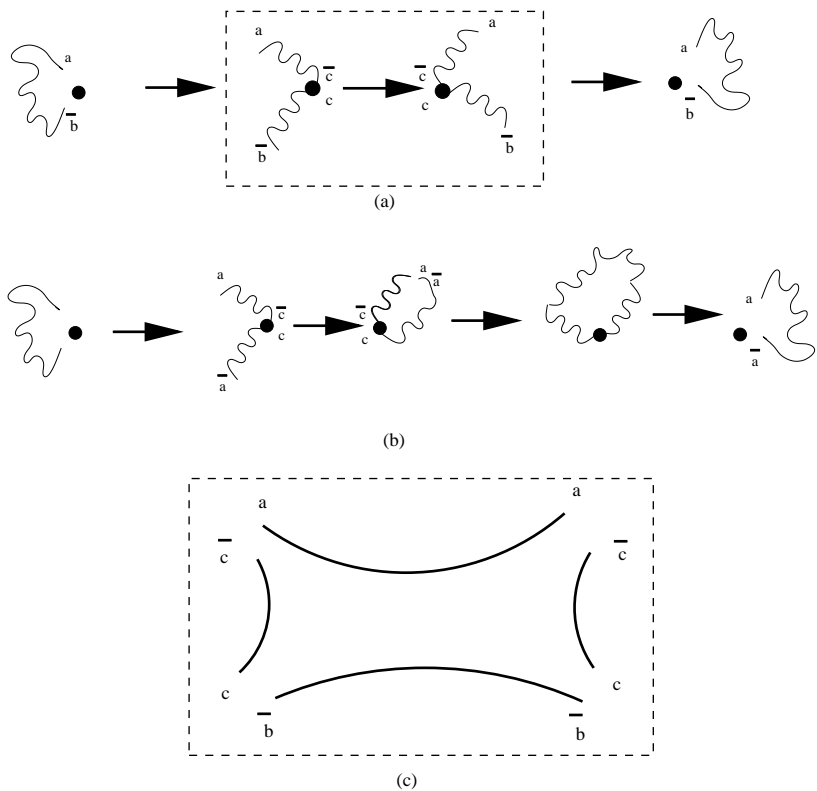}
\caption{\textbf{(a):} the splitting/capture/re-emission process of a (generic)
matter string by a D-particle from a target-space point of view. \textbf{(b)} The
same process but for photons (or in general particles in the Cartan subalgebra of
the gauge group in this (intersecting) brane world scenario.
\textbf{(c):} The four-point string scattering amplitude (corresponding to the parts
inside the dashed box of (a)) between the constituent open ND strings of the splitting
process. Latin indices at the end-points of the open string refer to the
brane worlds these strings are attached to.}
\label{fig:scatt}
\end{figure}

We denote a generic SM particle as an open string $a\bar b$ with both ends
on the D7-branes. For $a=b$,
we obtain the gauge fields related to the
Cartan subalgebras of the SM gauge groups, and their supersymmetric
partners (gauginos), for example, the
photon, $Z^0$ gauge boson and the gluons associated with the $\lambda^{3}$
and $\lambda^{8} $ Gell-Mann matrices of the $SU(3)_C$ group.
For $a\not= b$, we obtain the other particles, for example, the electron,
neutrinos, and $W^{\pm}_{\mu}$ boson, etc.
As in Fig. \ref{fig:scatt}(a), when the open string
$a \bar b$ passes through the  D3-brane, it can be split and become
two open strings (corresponding to the ND particles)
 $a \bar c$ and $c \bar b$ with one end on the D7-brane
($a$ or $\bar b$) and one end on the D3-brane ($c$ or $\bar c$).
Then, we can have the
 two to two process and have two out-going particles
arising from the open strings  $a \bar c$ and $c \bar b$. Finally,
we can have an out going particle denoted as open string $a\bar b$.
In particular, for $a=b$, we can have $s$-channel process  at
the leading order, and
plot it in the Fig. \ref{fig:scatt}(b). Interestingly,
the time delays arise from the two to two process in the box of
 Fig. \ref{fig:scatt}(a), and we plot the corresponding string diagram in Fig. \ref{fig:scatt}(c).

To calculate the time delays, we consider the four-fermion scattering
amplitude and use the results in Ref.~\cite{benakli} for simplicity.
We can discuss the other scattering amplitudes similarly, for example,
the four-scalar scattering amplitude, and the results are the same.
The total four-fermion scattering amplitude is obtained by summing up
 the various orderings~\cite{benakli}:
\begin{eqnarray}
\mathcal{A}_{\rm total } \equiv \mathcal{A}(1,2,3,4)
+ \mathcal{A}(1,3,2,4) +
 \mathcal{A}(1,2,4,3)~,~\,
\end{eqnarray}
where
\begin{eqnarray}
 \mathcal{A}(1,2,3,4) \equiv A(1,2,3,4) + A(4,3,2,1)~.~\,
\end{eqnarray}
$A(1,2,3,4)$ is the standard four-point ordered scattering amplitude
\begin{eqnarray}
&& \! \!(\!2\pi\!)^4 \! \delta^{(4)}(\sum_a k_a)\! A(1,2,3,4)\!
=  \! \frac{-i}{g_s l_s^4}\!
\int_{0}^{1}\!\! dx\!
\nonumber \\
&&
\left<\! {\cal V}^{(1)} (0,\!k_1) {\cal
V}^{(2)} (x,\!k_2) {\cal V}^{(3)} (1,\!k_3) {\cal V}^{(4)} (\infty,\!k_4)\!\right> ~,~\,
\end{eqnarray}
where $k_i$ are the space-time momenta, and we used the SL(2,R)
symmetry to fix three out of the four $x_i$ positions on the boundary of the upper
half plane, representing the
insertions of the open string vertex fermionic ND operators ${\cal V}^{(i)}$,
$i=1,\dots 4$, defined appropriately in Ref.~\cite{benakli}, describing the emission
of a massless fermion originating from a string stretched between the D7 brane
and the D3 brane.

The amplitudes depend on kinematical invariants expressible in terms of the
Mandelstam variables: $s=-(k_1 + k_2)^2$, $t=-(k_2 + k_3)^2$ and $u=-(k_1 + k_3)^2$,
for which $s + t + u =0$ for massless particles.
The ordered four-point amplitude $\mathcal{A}(1,2,3,4)$ is given by
\begin{eqnarray}
&& \mathcal{A} (1_{j_1 I_1},2_{j_2 I_2},3_{j_3 I_3},4_{j_4 I_4})= \nonumber \\
&&- { g_s} l_s^2 \int_0^1 dx \, \, x^{-1 -s\, l_s^2}\, \, \,
(1-x)^{-1 -t\, l_s^2} \, \, \,  \frac {1}{ [F (x)]^2 } \, \times  \nonumber \\
 &&  \left[   {\bar u}^{(1)} \gamma_{\mu} u^{(2)}
{\bar u}^{(4)} \gamma^{\mu} u^{(3)} (1-x) + {\bar u}^{(1)} \gamma_{\mu}
u^{(4)} {\bar u}^{(2)} \gamma^{\mu} u^{(3)}  x \right ] \,  \nonumber \\
&&  \times \{ \eta
\delta_{I_1,{\bar I_2}} \delta_{I_3,{\bar I_4}}
\delta_{{\bar j_1}, j_4} \delta_{j_2,{\bar j_3}}
\sum_{m\in {\bf Z}}  \, \,
e^{ - {\pi} {\tau}\,
   m ^2 \, \ell_s^2 /R^{\prime 2}   }
\nonumber \\
&& +  \delta_{j_1,{\bar j_2}}
\delta_{j_3,{\bar j_4}}
\delta_{{\bar I_1}, I_4} \delta_{I_2,{\bar I_3}}
\sum_{n\in {\bf Z}}  e^{-  {\pi \tau}   n^2
\,  R^{\prime 2} / \ell_s^{2} } \}~,~\,
\label{4ampl}
\end{eqnarray}
where $F(x)\equiv F(1/2; 1/2; 1; x)$ is the hypergeometric function,
$\tau (x) = F(1-x)/F(x)$,
$j_i$ and $I_i$ with $i=1, ~2, ~3, ~4$
are indices on the D7-branes and D3-branes, respectively, and $\eta$ is
\begin{eqnarray}
\eta={{(1.55\ell_s)^4} \over {V_{A3} R'}}~,~\,
\label{ETA-P}
\end{eqnarray}
in the notation of \cite{benakli},  $u$ is a fermion polarization spinor,
 and the dependence of the
appropriate Chan-Paton factors has been made explicit. Thus,
taking $F(x)\simeq 1$ we obtain
\begin{eqnarray}
\label{4ampldetail}
&&\mathcal{A}(1,2,3,4) \propto  g_s\ell_s^2 \left( t\ell_s^2
{\overline u}^{(1)}\gamma_\mu u^{(2)}{\overline u}^{(4)}\gamma^\mu u^{(3)}
\right. \nonumber\\ && \left.
+ s\ell_s^2{\overline u}^{(1)}\gamma_\mu u^{(4)}{\overline u}^{(2)}
\gamma^\mu u^{(3)}\right)
 \times \frac{\Gamma(-s\ell_s^2)\Gamma(-t\ell_s^2)}{\Gamma(1 + u\ell_s^2)}~,
\nonumber \\
&&\mathcal{A}(1,3,2,4) \propto
 g_s\ell_s^2 \left( t\ell_s^2
{\overline u}^{(1)}\gamma_\mu u^{(3)}{\overline u}^{(4)}\gamma^\mu u^{(2)}
\right. \nonumber\\ && \left.
+ u\ell_s^2{\overline u}^{(1)}\gamma_\mu u^{(4)}{\overline u}^{(3)}
\gamma^\mu u^{(2)}\right)  \times
\frac{\Gamma(-u\ell_s^2)\Gamma(-t\ell_s^2)}{\Gamma(1 + s\ell_s^2)}~, \nonumber \\
&&\mathcal{A}(1,2,4,3) \propto  g_s\ell_s^2 \left( u\ell_s^2
{\overline u}^{(1)}\gamma_\mu u^{(2)}{\overline u}^{(3)}\gamma^\mu u^{(4)}
\right. \nonumber\\ && \left.
+ s\ell_s^2{\overline u}^{(1)}\gamma_\mu u^{(3)}{\overline u}^{(2)}
\gamma^\mu u^{(4)}\right)
 \times \frac{\Gamma(-s\ell_s^2)\Gamma(-u\ell_s^2)}{\Gamma(1 + t\ell_s^2)}~,~\,
\end{eqnarray}
where the proportionality symbols incorporate Kaluza-Klein
or winding mode contributions,
which do not contribute to the time delays. Technically, it should be noted that the novelty of our results above, as compared with those of \cite{benakli}, lies on the specific compactification procedure we adopted, and the existence of a uniformly distributed population of D-particles (foam), leading to (\ref{coupl-N}).

Similarly to the discussion in Ref.~\cite{sussk}, time delays arise from  the
amplitude $\mathcal{A}(1,2,3,4)$ by considering
backward scattering $u=0$. Noting that $s+t+u=0$ for massless particles,
the first term in $\mathcal{A}(1,2,3,4)$
in Eq. (\ref{4ampldetail})  for $u=0$ is proportional to
\begin{eqnarray}
t\ell_s^2 \Gamma(-s\ell_s^2) \Gamma(-t \ell_s^2) &=& -s \ell_s^2 \Gamma(-s\ell_s^2)
\Gamma(s \ell_s^2)
\nonumber\\ &=&
{{\pi} \over {\sin(\pi s\ell_s^2)}}~.~\,
\end{eqnarray}
It has poles at $s=n/\ell_s^2$. The divergence of the amplitude at the
poles is an essential physical feature of the amplitude,
a resonance corresponding
to the propagation of an intermediate string state over long space-time
distances. To define the poles we use the correct $\epsilon$ prescription
replacing $s \to s+ i\epsilon$, which shift the poles off the real axis.
Thus, the functions $1/\sin(\pi s\ell_s^2)$ can be expanded as a power
series in $y$ which is
\begin{eqnarray}
y~=~ e^{i\pi s \ell_s^2 - \epsilon} ~.~\,
\end{eqnarray}
On noting that $s = E^2$, we obtain the time delay at the lowest order
\begin{eqnarray}
\Delta t = E \ell_s^2~.~\,
\label{t-delay}
\end{eqnarray}

Let us discuss the time delays at leading order
for concrete particles.
We will assume that $\eta $ is a small number about
$10^{-6}$ or smaller. Then, for the gauge fields (and their corresponding
gauginos) which are
related to the Cartan subalgebras of the SM gauge groups,
all the amplitudes $\mathcal{A}(1,2,3,4)$, $\mathcal{A}(1,3, 2,4)$,
and  $\mathcal{A}(1,2,4,3)$ will give the dominant contributions
to the total amplitude due to $j_1={\bar j_2}$. Thus,
they will have time delays as given
in Eq. (\ref{t-delay}). The resulting delay for photon
is independent of its polarization, and thus there
is \emph{no birefringence}, thereby leading to the evasion
of the relevant stringent astrophysical constraints~\cite{uv,grb,macio}.

However, for the other particles, we have $j_1\not= {\bar j_2}$, and
then only the amplitude $\mathcal{A}(1,3, 2,4)$ gives dominant
contribution. Considering backward scattering~\cite{sussk} $u=0$ and
$s+t+u=0$, we obtain
\begin{eqnarray}
\mathcal{A}(1,3,2,4) & \propto &
 g_s\ell_s^2 \left({1\over {u\ell_s^2}}
{\overline u}^{(1)}\gamma_\mu u^{(3)}{\overline u}^{(4)}\gamma^\mu u^{(2)}
\right. \nonumber\\ && \left.
- {1\over {s\ell_s^2}}
u\ell_s^2{\overline u}^{(1)}\gamma_\mu u^{(4)}{\overline u}^{(3)}
\gamma^\mu u^{(2)}\right)  ~.~\,
\end{eqnarray}
Because they are just the pole terms,  we do not have time delays for
 other particles with $j_1\not= {\bar j_2}$ at the leading order, for example,
$W^{\pm}_{\mu}$ boson, electron, and neutrinos, etc.
At order $\eta$ (${\cal O} (\eta)$), we have time delays for
these particles, which arise from the forth line in Eq. (\ref{4ampl}).
The dispersion relation for the electron can be parametrised as
follows
\begin{eqnarray}
E^2 ~=~ p^2 +m_e^2 - \eta p^3/M_{\rm St}~,~\,
\label{Elec-Disp}
\end{eqnarray}
where $M_{St}$ is about the string scale.
From the Crab Nebula synchrotron radiation observations, we obtain the constraint on
$\eta$~\cite{crab2}
\begin{eqnarray}
\eta  \leq 10^{-6}~.~\,
\end{eqnarray}
This can be realized easily, for example, by taking the following
$V_{A3}$ and $R'$ in Eq. (\ref{ETA-P})
\begin{eqnarray}
V_{A3} \sim (10 \ell_s)^3~,~~~R'\sim 338 \ell_s~.~\,
\end{eqnarray}

Interestingly, because the neutrinos have a similar
dispersion relation as the electron in Eq. (\ref{Elec-Disp}),
we might have implications in neutrino (oscillation) physics,
and this may lead to important phenomenological constraints~\cite{sarkar, foffa}
which will be studied elsewhere.

Finally, we close this section by mentioning that the inclusion of the recoil
motion of the D-particles (wrapped up D3-branes in the current model) is
straightforward and parallels the discussion leading to
Eq.~(\ref{timerecoil}) in the introduction, at least in the limit we are considering here,
namely the case where the radii of the compactified D3-branes are not much larger
than the string length in the problem, the latter being assumed to correspond to conventional large string mass scales of order $10^{17}-10^{18}$~GeV. For such scales,
the D3 brane is effectively viewed as point like to a good approximation,
for all practical (low-energy phenomenology) purposes. Such a recoil, then,
will lead to corrections suppressed by quadratic and higher-order powers
of the corresponding quantum gravity scale, which in this case will be the
effective D-particle (wrapped up D3-brane) mass given in Eq.~(\ref{D3-Mass}).
As already pointed out, the D-particle mass can be much heavier
than the string scale, upon choosing compactification radii larger than the string length,
and in such heavy D-brane cases recoil effects can be easily neglected.

It should also be noticed that the presence of recoil leads to an
effective string coupling $g_s^{\rm eff}$ in Eq.~(\ref{effstringcoupl}), which implies that
for relativistic scattering of D-particles, {\it i.e.}, when the recoil velocity
is of order of the speed of light, the corresponding string scattering
amplitudes, from which the delays are evaluated, will be suppressed.
Thus, although there would be infinitely long delays practically for
such relativistic particles (c.f. (\ref{timerecoil})), nevertheless the pertinent amplitudes
would vanish, as in such a case they are given by expressions of the form (\ref{4ampl})
but with the factors $g_s$ being replaced by the effective string coupling  $g_s^{\rm eff}$.
Hence, there would be no observable effects, since
the corresponding cross section would be vanishing~\footnote{For high-energy (compared to the string mass scale) string scattering, the higher genus world-sheet amplitudes are more relevant. Although a resummation of higher world-sheet topologies is not possible in a first quantised string framework, and hence exact expressions for fully quantum high-energy string amplitudes are not available, nevertheless time delays for a genus $G$ amplitude have been calculated in \cite{ray} and found suppressed by a factor $1/(G+1)$, compared to the genus G=0 tree level amplitudes, considered in \cite{sussk} and here.}.

This observation may have important consequences when one considers
models of string D-foam with low string
scales $M_s$, say of a few TeV. Indeed, in our type IIB foam, where each D-particle is viewed
as a compactified D3-brane, wrapped around three-cycles with very
small radii, so that one has an effectively point-like behaviour at low energies,
the corresponding mass scales can be of order of the string scale
$M_s$, provided  the compactification radii are of order of the string length, $R_i \sim \ell_s$ and $g_s\sim 1$.
Thus, incident particles with energies higher than the order-TeV string scale $M_s$ on such D-branes can easily
induce recoil velocities close to the speed of light, and, as a consequence of the above discussion,
for such energetic particles the foam would look transparent~\footnote{Similar observations are valid for
the type IA (or IIA)  D-foam model of \cite{emnnewuncert}, involving truly point-like D-particles of mass $M_s/g_s$.}.
On the other hand, for large compactification radii compared to the string length
 $R_i \gg 10 \ell_s$, the resulting D-particle masses can be very large (c.f. (\ref{D3-Mass})), thereby alleviating the transparency of the D-foam to highly energetic particles in such models.

 What this means is that, in the context of the above model, for low-string scales, say $M_s ={\cal O}$(TeV), time delays should be expected for photons with energies much lower than TeV scales, whilst photons with energies above TeV will essentially remain undisturbed by the presence of D-particles, thereby not exhibiting any appreciable foam-induced time delays. Therefore, astrophysical observations using cosmic photons, such as those by MAGIC., H.E.S.S. and FERMI Telescopes, can be used to discriminate low- from high-string scale models, provided of course that the respective source mechanisms for the production of the cosmic photons are understood.

  These remarks are useful to bear in mind when considering phenomenological constraints on D-foam models imposed by observations of very high energetic cosmic particles (with energies of order (or higher than) $10^{20}$ eV)~\cite{sigl}.

\section{Conclusions}

In this work we considered Type IIB string theory with D3-branes and
D7-branes. By wrapping up the D3 branes around three cycles, and
 the D7 branes around four cycles, appropriately, we have focused on
a representation of a space-time foam situation
in this type of string theory, which extends non-trivially the type-IA (and  IIA ) string
theory foam model of \cite{emnnewuncert} to the phenomenologically
richer type IIB string theory.

We derived the vacuum refractive index for photon, and found that it
depends linearly on the photon energy, as in the case of \cite{emnnewuncert}. However,
in the current model of foam, there are also non-trivial refractive indices for charged probes,
such as electrons. This was consistent with electric charge conservation, as a result of our compactification procedures, regarding the representation of D-particles as wrapped up D3-branes, as well as our specific ``foam'' aspects of the model, namely the presence of a uniform population of such D-particles (leading to non trivial couplings (\ref{coupl-N})).
Nevertheless, as a result of appropriate compactification, the time-delay
for electron can be small enough to satisfy the  stringent
constraints coming from synchrotron radiation of the Crab Nebula.

A final remark we would like to make concerns the cosmology of the model, which
would epitomize the phenomenology of such space-time foam models. The constraint
on the density of the D3(compactified)-brane defects at the current epoch of the
D7(compactified)-brane Universe, representing our world, in order to fit the photon
time-delay data available to date~\cite{MAGIC2,hessnew,grbglast} needs to be taken
into account in conjunction with
the other cosmological parameters. This would affect the energy budget of
the Universe, and through this, the relevant astro-particle phenomenology,
in similar spirit
to the type-IIA (and type IA) D-particle foam model~\cite{emnw,emnnewuncert}.
However, the calculation of the various cosmological parameters
and in general the evolution of such a Universe depends highly on the model,
 for instance on details of the target-space supersymmetry breaking after
compactification, the bulk physics, etc.
The study of such issues is left for the future.

\section*{Acknowledgements}

This research was supported in part by the
Cambridge-Mitchell Collaboration in Theoretical Cosmology (TL),
by the Natural Science Foundation of China under grant No. 10821504 (TL),
by the Chinese Academy of Sciences under the grant No. KJCX3-SYW-N2 (TL),
by the European Union through the Marie Curie Research and Training Network \emph{UniverseNet}
(MRTN-2006-035863) (NEM), and by
DOE grant DE-FG02-95ER40917 (DVN).

\end{document}